\pdfoutput=1
\documentclass[reprint,
 amsmath,amssymb,
 aps,
]{revtex4-2}

\usepackage{graphicx}
\usepackage{dcolumn}
\usepackage{bm}
\usepackage{hyperref}
\usepackage{booktabs,tabularx}
\usepackage{multirow}
\usepackage{booktabs, threeparttable}
\usepackage{array}
\usepackage{siunitx}
\usepackage{xparse}
\usepackage{caption}

\usepackage{orcidlink}

\DeclareUnicodeCharacter{2212}{\ensuremath{-}}

\hypersetup{
    colorlinks=true,
    linkcolor=magenta,
    filecolor=magenta,      
    urlcolor=blue,
    citecolor=blue,
    }

\begin{document}

\title{Observation of partonic collectivity via $p_{\rm T}$-differential radial flow fluctuations in Au+Au collisions at $\sqrt{s_{\rm NN}} = 200$ GeV}

\author{Rohit Agarwala$^{1}$ \orcidlink{0009-0009-8413-1123}}
\email{therohitagarwala@buniv.edu.in}
\author{Dipankar Basak$^{1,2}$ \orcidlink{0000-0002-7878-6256}}
\email{dipankar0001@gmail.com}
\author{Kalyan Dey$^{1}$ \orcidlink{0000-0002-4633-2946}}
\email{kalyan.dey@buniv.edu.in (corresponding author)}
\affiliation{$^{1}$Department of Physics, Bodoland University, Debargaon, Kokrajhar - 783370 Assam, India}
\affiliation{$^{2}$Department of Physics, Kokrajhar University, Kokrajhar - 783370 Assam, India}

\date{\today}

\begin{abstract}
We report the observation of partonic radial collectivity in Au+Au collisions at $\sqrt{s_{\rm NN}} = 200$~GeV via the $p_{\rm T}$-differential flow observable $v_{0}(p_{\rm T})$ using the \texttt{AMPT} String Melting model. For inclusive charged hadrons, we establish three signatures of collectivity: long-range pseudorapidity correlations, the factorization of two-particle correlations, and a centrality-independent scaling of $v_{0}(p_{\rm T})$ normalized by its $p_{\rm T}$-integrated value $v_{0}$, analogous to anisotropic flow. For identified particles ($\pi^{\pm}, K^{\pm}, p + \overline p$), the $v_{0}(p_{\rm T})$ spectra show mass ordering at low-$p_{\rm T}$ and meson-baryon separation at intermediate-$p_{\rm T}$. In \textit{central} collisions, $v_{0}(p_{\rm T})/n_{q}$ exhibits robust \textit{Number of Constituent Quark} (NCQ) scaling with $(m_{\rm T} - m_{0})/n_{q}$, a scaling that breaks down in \textit{peripheral} collisions and is more precise at RHIC than at LHC energies, consistent with earlier $v_{2}$ studies. These findings provide strong evidence that radial collectivity originates predominantly at the partonic stage, extending the paradigm of quark-level dynamics from anisotropic to isotropic flow.
\end{abstract}

\maketitle

\section{Introduction}
\label{introduction}
Ultra-relativistic nucleus-nucleus collisions provide access to hot and dense QCD matter, commonly known as the quark-gluon plasma (QGP), where quarks are no longer confined within hadrons~\cite{shuryak1980,mclerran1986,bbback2005phobos,adams2005star,adcox2005phenix}.
The formation of this strongly interacting plasma is now supported by a broad set of experimental observations. Two of its most compelling signatures are jet quenching~\cite{gyulassy1990,qin2015} and the \textit{Number of Constituent Quark} (NCQ) scaling of the second-order flow harmonic $v_{2}$~\cite{adare2007phenix,zhu2025violation}. The observed scaling of the elliptic flow of hadrons collectively follows a universal curve when scaled by their number of constituent quarks, providing direct evidence for the emergence of partonic collectivity prior to hadronization. \\
\par
Complementary to elliptic flow, the isotropic radial flow originates from the strong pressure gradients generated at the early stage of the collision, driving the collective transverse expansion of the medium and providing insight into the equation of state and space-time evolution of the QGP~\cite{acharya2026alice}. Traditionally, radial flow has been characterized using a transverse momentum $p_{\rm T}$-integrated observable, obtained by performing simultaneous blast-wave fits to the $p_{\rm T}$-spectra of identified hadrons. Recently, a new observable, $v_{0}(p_{\mathrm{T}})$, has been introduced to quantify radial flow through the correlation between the particle yield in a given $p_{\mathrm{T}}$-bin and the eventwise mean transverse momentum~\cite{schenke2020,parida2024}. This differential measure provides a more direct handle on collective expansion and exhibits improved sensitivity compared to traditional blast-wave-based extractions that rely on model-dependent assumptions~\cite{tang2009,saha2025}. Recent measurements of $v_{0}(p_{\mathrm{T}})$ by the ALICE Collaboration~\cite{acharya2026alice} show a clear mass-ordering at low-$p_{\mathrm{T}}$ and a meson-baryon splitting at intermediate-$p_{\mathrm{T}}$, along with an approximate NCQ scaling behavior. These features are qualitatively consistent with the systematics previously observed in $v_{2}$ measurements~\cite{lin2002prl,adamczyk2016star,adams2005multistrange,
adare2007phenix, singha2016}, where similar mass-ordering and scaling patterns have been interpreted as signatures of collective expansion 
and partonic dynamics. Further, the observable is also sensitive to the transport properties of the system, with a pronounced dependence on the bulk viscosity~\cite{acharya2026alice,du2026}, in contrast to elliptic flow, whose magnitude is influenced by both bulk and shear viscous contributions~\cite{song2010,denicol2010}. Inclusive charged-particle measurements by ATLAS collaboration~\cite{aad2026atlas} in Pb+Pb collisions demonstrate that the radial-flow observable $v_{0}(p_{\mathrm{T}})$ exhibits genuine collective behavior, manifested through long-range pseudorapidity correlations, factorization of two-particle correlations into single-particle $v_{0}(p_{\mathrm{T}})$, and a nearly centrality-independent normalized shape $v_{0}(p_{\mathrm{T}})/v_{0}$ at low-$p_{\mathrm{T}}$
\cite{aad2026atlas,bhatta2025disentangling}. Further, as reported in Refs.~\cite{wan2025tracing,saha2025}, the hydrodynamic-inspired blast-wave model with fluctuating kinetic freeze-out temperature and flow velocity can explain the mass-ordering of $v_{0}(p_{\rm T})$ for identified charged hadrons at low-$p_{\rm T}$. Moreover, as an extension, an \texttt{AMPT}-based study~\cite{wan2025tracing} with string melting configuration demonstrated that the meson-baryon crossing at intermediate-$p_{\rm T}$ originates primarily from the quark coalescence mechanism. \\
\par
Crucially, the three hallmarks of collectivity and NCQ scaling of $v_{0}(p_{\rm T})$ have thus far been established at LHC energy (Pb+Pb, 5.02 TeV). Whether these signatures persist at RHIC energies remains an open question. In this work, we examine the corresponding factorization properties and scaling behaviors, including the NCQ scaling of $v_{0}(p_{\rm T})$, in Au+Au collisions at $\sqrt{s_{\rm NN}}=200$ GeV within the \texttt{AMPT} string-melting framework. Extending these tests to lower collision energies enables an assessment of the energy dependence of the observed collective patterns and provides insight into the robustness of the scaling features reported at the LHC. \\
\par
The paper is organized as follows. Section~\ref{sec:2} outlines the \texttt{AMPT} and \texttt{PYTHIA8/Angantyr} event generators and the specific configurations used in this study. Section~\ref{sec:3} describes the event selection criteria and the methodology for calculating $v_{0}(p_{\mathrm{T}})$ and its integrated form $v_{0}$. Section~\ref{sec:4} presents the results, including $v_{0}(p_{\mathrm{T}})$ for inclusive charged hadrons and identified particles. Finally, Section~\ref{sec:5} summarizes the main findings.
\section{Model Description} \label{sec:2}

\subsection{The \texttt{AMPT} model}

The multi-phase transport (\texttt{AMPT}) model offers a comprehensive microscopic description of heavy-ion collisions, comprising four distinct stages~\cite{lin2005}. In the initial stage, \texttt{HIJING}~\cite{wang1991hijing,gyulassy1994hijing} generates the fluctuating nucleon-nucleon interactions, including hard and semi-hard scatterings, jet production, nuclear shadowing, and the associated excited strings. In the default version, these strings hadronize via Lund string fragmentation, whereas in the string-melting (SM) version, they are first converted to their constituent quarks and antiquarks according to the endpoint flavors and momenta, providing a more suitable setup for studying partonic collectivity. The subsequent partonic evolution (present only in the SM scenario) is modeled with Zhang’s Parton Cascade (\texttt{ZPC})~\cite{zhang1998}, which includes elastic two-body scatterings among quarks and gluons with parameterized cross-sections, followed by hadronization through a quark coalescence prescription in which nearby partons recombine into mesons and baryons.  Finally, the hadronic phase is described by the \texttt{ART} (A Relativistic Transport) model~\cite{baoanli1995}, which propagates the produced hadrons through elastic and inelastic meson–meson, meson–baryon, and baryon-baryon interactions, including resonance formation and decay channels. For the present study, we employ the \texttt{AMPT} model \texttt{version 2.26t9b}.
\subsection{The \texttt{PYTHIA8/Angantyr} model}

The \texttt{Angantyr} model~\cite{bierlich2018angantyr} extends the \texttt{PYTHIA8} framework~\cite{bierlich2022} to simulate $p$A and AA collisions by superimposing multiple fluctuating nucleon–nucleon sub-collisions within the standard $pp$ formalism, without introducing an explicit hydrodynamic medium or QGP phase. The nuclear geometry is determined via a Glauber-model calculation that includes event-by-event fluctuations of the nucleon–nucleon cross section. Each binary encounter is classified as non-diffractive, single-diffractive, or double-diffractive excitation following a wounded-nucleon-inspired implementation of the 
Fritiof model~\cite{andersson1987fritiof}. Multi-parton interactions and color reconnection are treated within the underlying \texttt{PYTHIA8} framework, enabling the generation of fully exclusive hadronic final states and a realistic description of global observables such as charged-particle multiplicity and transverse-momentum spectra in heavy-ion collisions~\cite{vieira2019,bierlich2021,ortiz2024}. In this work, \texttt{PYTHIA8/Angantyr} is employed as a non-collective baseline, where particle production arises from superimposed nucleon–nucleon interactions without medium-induced collective expansion. Pb+Pb collision events at $\sqrt{s_{\rm NN}}=5.02$ TeV are generated using \texttt{version 8.316} of the model.
\begin{figure}
    \begin{center}
    \includegraphics[width=\linewidth]{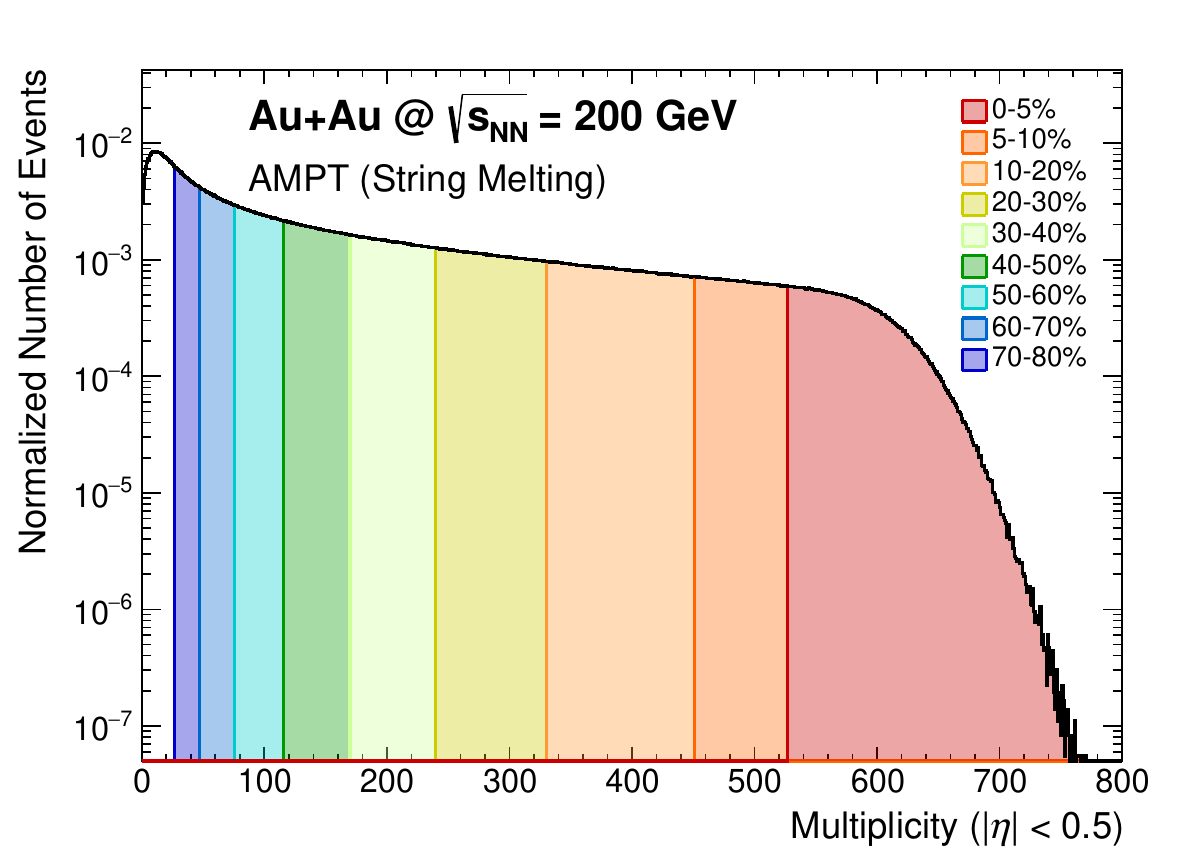}
    \includegraphics[width=\linewidth]{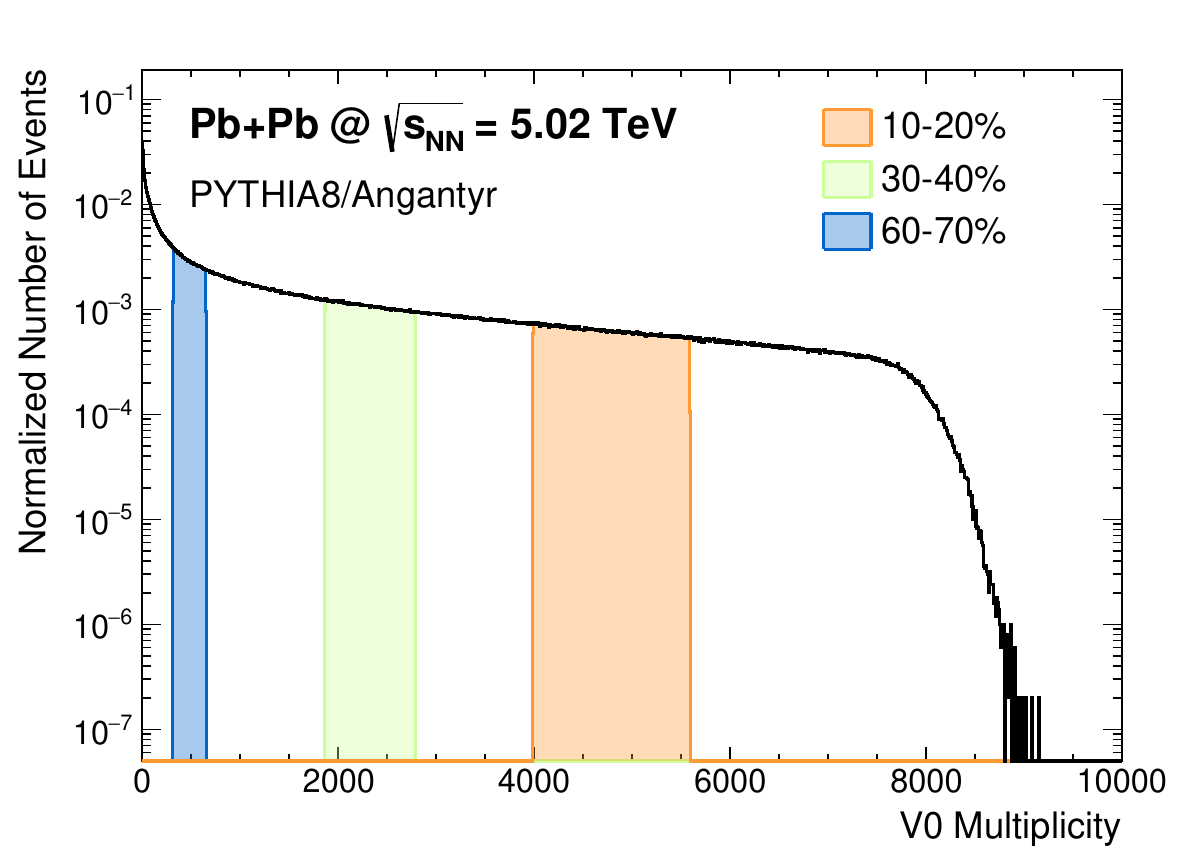}
    \end{center}
    \caption{(Color online) Multiplicity distribution in Au+Au collisions at $\sqrt{s_{\rm NN}} = 200$ GeV (\textit{top}); and Pb+Pb collisions at $\sqrt{s_{\rm NN}} = 5.02$ TeV (\textit{bottom}), simulated using \texttt{AMPT} String Melting (SM) and \texttt{PYTHIA8/Angantyr} models respectively, with multiple centrality classes from \textit{most central} to \textit{peripheral}. The black curve represents the total number of events analyzed.}
    \label{fig:AuAu_PbPb_multdist}
\end{figure}
\section{Methodology} \label{sec:3}

\subsection{Event Selection and Centrality Determination}

We analyze Au+Au collisions at $\sqrt{s_{\rm NN}}=200$ GeV and Pb+Pb collisions at $\sqrt{s_{\rm NN}}=5.02$ TeV. For Au+Au, 36.31 million minimum-bias events are generated using the \texttt{AMPT} model in the string-melting configuration. For Pb+Pb, 5 million minimum-bias events are simulated with \texttt{PYTHIA8/Angantyr} to provide a non-collective baseline. In addition, published ALICE results for Pb+Pb collisions at $\sqrt{s_{\rm NN}}=5.02$ TeV~\cite{acharya2026alice} are used for comparison. For Au+Au collisions, centrality is defined via the charged-particle multiplicity at midrapidity ($|\eta|\leq0.5$). For Pb+Pb collisions, 
centrality is determined using charged-particle multiplicity in the V0 detector acceptance ($2.8<\eta<5.1$ and $-3.7<\eta<-1.7$), corresponding to the V0M estimator employed by ALICE~\cite{acharya2026alice}. The multiplicity distributions are shown in Fig.~\ref{fig:AuAu_PbPb_multdist}, and the corresponding multiplicity intervals for each centrality class are summarized in Tab.~\ref{tab:multiplicity}.
\begin{table}[htbp]
\centering
\caption{Multiplicity ranges ($dN_{\rm ch}/d\eta$) for different centrality intervals in Au+Au collisions at $\sqrt{s_{\rm NN}} = 200$ GeV (mid-rapidity centrality, \texttt{AMPT-SM}) and Pb+Pb collisions at $\sqrt{s_{\rm NN}} = 5.02$ TeV (V0M centrality, \texttt{PYTHIA8/Angantyr}).}
\label{tab:multiplicity}
\begin{tabular}{cc}
\toprule\toprule
\multicolumn{2}{c}{Au+Au, $\sqrt{s_{\rm NN}} = 200$ GeV (\texttt{AMPT-SM})} \\
\cmidrule(lr){1-2}
Centrality (\%) & $dN_{\rm ch}/d\eta$ \\
\midrule
$0$--$5$   & $527$--$793$  \\
$5$--$10$  & $451$--$527$  \\
$10$--$20$ & $330$--$451$  \\
$20$--$30$ & $240$--$330$  \\
$30$--$40$ & $170$--$240$  \\
$40$--$50$ & $116$--$170$  \\
$50$--$60$ & $76$--$116$   \\
$60$--$70$ & $47$--$76$    \\
$70$--$80$ & $27$--$47$    \\
\midrule\midrule
\multicolumn{2}{c}{Pb+Pb, $\sqrt{s_{\rm NN}} = 5.02$ TeV (\texttt{Angantyr})} \\
\cmidrule(lr){1-2}
Centrality (\%) & $dN_{\rm ch}/d\eta$ \\
\midrule
$10$--$20$ & $3992$--$5591$ \\
$30$--$40$ & $1862$--$2788$ \\
$60$--$70$ & $317$--$653$   \\
\bottomrule\bottomrule
\end{tabular}
\end{table}

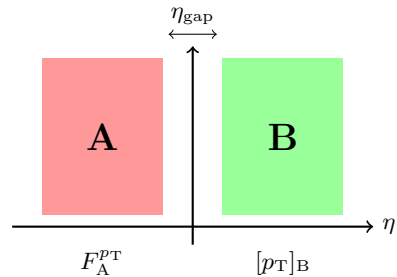
\begin{figure}
    \centering
    \begin{tikzpicture}[scale=0.8]
        \draw[thick,->] (-3,0) -- (3,0) node[right] {$\eta$};
        \draw[thick,->] (0,-0.3) -- (0,3) node[above]{} ;
        
        \fill[red!40] (-2.5,0.2) rectangle (-0.5,2.8);
        \node at (-1.5,1.5) {\Large \textbf{A}};
        
        \fill[green!40] (0.5,0.2) rectangle (2.5,2.8);
        \node at (1.5,1.5) {\Large \textbf{B}};
        
        \draw[<->] (-0.4,3.2) -- (0.4,3.2);
        \node at (0,3.5) {\small $\eta_{\rm gap}$};

        \node at (-1.5,-0.6) {$ F_{\rm A}^{p_{\rm T}}$};
        \node at (1.5,-0.6) {$[p_{\rm T}]_{\rm B}$};
    \end{tikzpicture}
    \caption{Cartoon illustrating the $\eta$ regions used in the analysis. Regions A (\textit{red}) and B (\textit{green}) are separated by a gap $\eta_{\rm gap}$. The observable $F_{\rm A}^{p_{\rm T}}$ is correlated with the transverse momentum $[p_{\rm T}]_{\rm B}$ estimated in the opposite region.}
    \label{fig:eta_regions}
\end{figure}

\begin{figure*}
    \begin{center}
    \includegraphics[width=\linewidth]{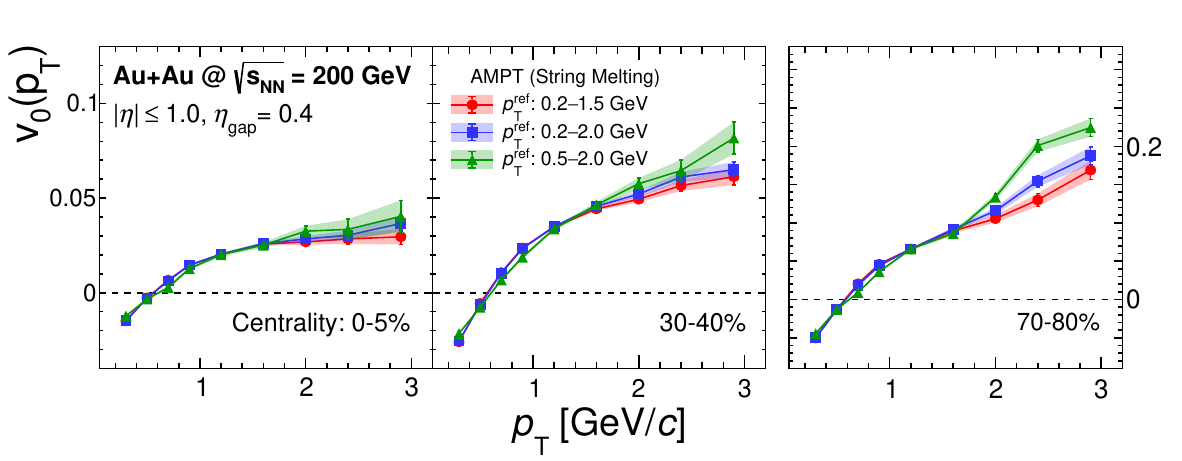}
    \end{center}
    \caption{(Color online) Transverse momentum dependence of the radial flow observable $v_{0}(p_{\rm T})$ plotted for three $p_{\rm T}^{\rm ref}$ selections at three centrality intervals: 0--5\% (\textit{most central}), 30--40\% (\textit{semi-central}) and 70--80\% (\textit{peripheral}), for Au+Au collisions simulated using \texttt{AMPT-SM} at $\sqrt{s_{\rm NN}} = 200$ GeV.}
    \label{fig:pTref_v0pT}
\end{figure*}

\subsection{Observable} \label{subsec:observable_and_scaling}

For the present investigation, radial flow is estimated using a novel observable, $v_{0}(p_{\rm T})$, which provides a unique probe of long-range collective dynamics by quantifying event-by-event correlations between fluctuations in charged-particle multiplicity and fluctuations in the average transverse momentum~\cite{schenke2020,parida2024}. The observable $v_{0}(p_{\rm T})$ is defined as,
\begin{equation} \label{eq:v0pT}
    v_{0}(p_{\rm T}) = \frac{\langle F_{\rm A}^{p_{\rm T}} {[p_{\rm T}]}_{\rm B} \rangle - \langle F_{\rm A}^{p_{\rm T}} \rangle {\langle [p_{\rm T}]}_{\rm B} \rangle}{\langle F_{\rm A}^{p_{\rm T}} \rangle \sigma_{[p_{\rm T}]}},
\end{equation}
where, $F_{\rm A}^{p_{\rm T}}$ is the fraction of particles in the transverse momentum bin $p_{\rm T}$ within the pseudorapidity window A and is given by,
    \begin{equation}
        F_{\rm A}^{p_{\rm T}} = \frac{N_{\rm ch}^{\rm A}(p_{\rm T})}{N_{\rm ch}^{\rm A}},
    \end{equation}
    where $N_{\rm ch}^{\rm A}(p_{\rm T})$ is the number of particles in the $p_{\rm T}$ bin and $N_{\rm ch}^{\rm A}$ is the total charged-particle multiplicity in window A. The second term in the numerator of Eq.~(\ref{eq:v0pT}) is the event-averaged mean transverse momentum, estimated in a separate, longitudinally-displaced window B (illustrated in Fig.~\ref{fig:eta_regions}), with a gap, $\eta_{\rm gap} > 0$  to suppress short-range correlations, is given by,
    \begin{equation}
        {[p_{\rm T}]}_{\rm B} = \frac{1}{N_{\rm ch}^{\rm B}} \sum_{i \in B} p_{\rm T_{i}}
    \end{equation}

The term $\sigma_{[p_{\rm T}]}$ in the denominator of Eq.~(\ref{eq:v0pT}) is the standard deviation of mean-$p_{\rm T}$ and can be calculated as follows,
\begin{equation}
    \sigma_{[p_{\rm T}]} = \sqrt{\langle {[p_{\rm T}]}_{\rm A} {[p_{\rm T}]}_{\rm B} \rangle - \langle {[p_{\rm T}]}_{\rm A} \rangle \langle {[p_{\rm T}]}_{\rm B} \rangle}.
\end{equation}

The integrated observable $v_{0}$ quantifies event-by-event fluctuations  of the mean transverse momentum $[p_{\rm T}]$. Following Ref.~\cite{aad2026atlas}, and based on the fluctuation formalism of Ref.~\cite{schenke2020}, it is defined as the relative root-mean-square of the fluctuations of the event-wise mean transverse momentum evaluated within a chosen reference transverse-momentum range $R$,
\begin{equation} \label{eq:5}
v_{0} \equiv 
\frac{\sqrt{\langle (\delta[p_{\rm T}])^2 \rangle_{R}}}
{\langle [p_{\rm T}] \rangle_{R}} ,
\end{equation}
where $\delta[p_{\rm T}] = [p_{\rm T}] - \langle [p_{\rm T}] \rangle_{R}$, 
and $\langle \cdots \rangle_{R}$ denotes averaging over events with particles restricted to the reference interval $R$. 

Under the assumption of collective dynamics, fluctuations of the single-particle spectrum $\delta n(p_{\rm T})$ are linearly correlated with fluctuations of $[p_{\rm T}]$, leading to an approximate factorization of the normalized covariance. This yields a sum rule relating the integrated observable to its transverse-momentum-differential counterpart $v_{0}(p_{\rm T})$,
\begin{equation}
v_{0} \sim
\frac{\int_{p_{\rm T} \in R} 
(p_{\rm T} - \langle [p_{\rm T}] \rangle_{R}) 
\langle n(p_{\rm T}) \rangle 
v_{0}(p_{\rm T}) \, dp_{\rm T}}
{\int_{p_{\rm T} \in R} 
p_{\rm T} \langle n(p_{\rm T}) \rangle \, dp_{\rm T}} ,
\end{equation}
where $\langle n(p_{\rm T}) \rangle$ denotes the event-averaged single-particle spectrum.

\begin{figure*}
    \centering
    \includegraphics[width=\linewidth]{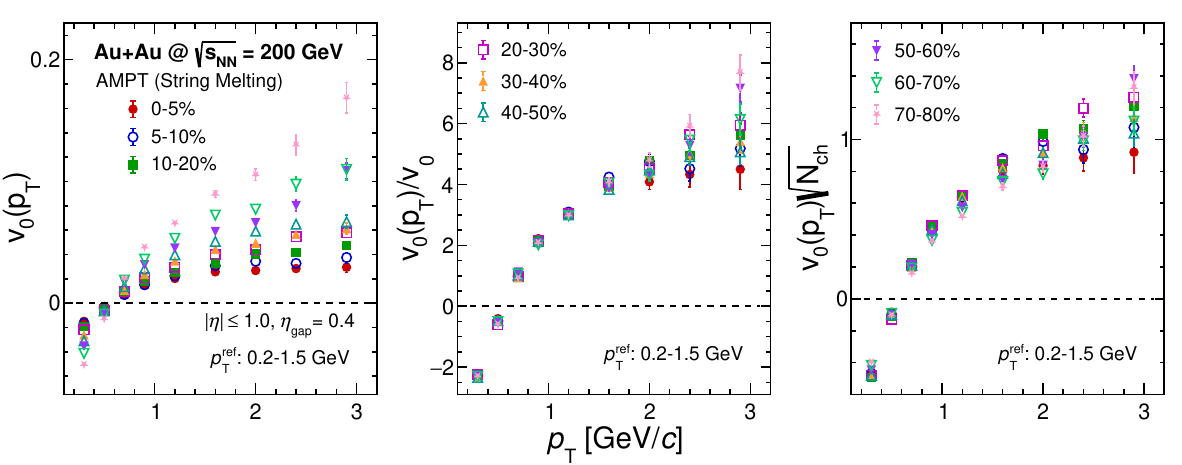}
    \caption{(Color online) Centrality dependence of $v_{0}(p_{\rm T})$ (\textit{left}), $v_{0}(p_{\rm T})/v_{0}$ (\textit{center}) and $v_{0}(p_{\rm T}) \sqrt{\rm N_{ch}}$ (\textit{right}), presented as a function of $p_{\rm T}$ for Au+Au simulations using \texttt{AMPT-SM} at $\sqrt{s_{\rm NN}} = 200$ GeV. A reference-$p_{\rm T}$ range of 0.2-1.5 GeV is selected for this particular analysis.}
    \label{fig:v0(pT)_v0(pT)/v0_v0(pT)sqrt{Nch}}
\end{figure*}

\section{Results and Discussion} \label{sec:4}
\subsection{Transverse momentum dependence of $v_{0}(p_{\rm T})$}

Fig.~\ref{fig:pTref_v0pT} shows the $p_{\rm T}$ dependence of the radial flow observable $v_{0}(p_{\rm T})$ for 0--5\% (\textit{most central}), 30--40\% (\textit{mid-central}), and 70--80\% (\textit{peripheral}) centrality classes in Au+Au collisions at $\sqrt{s_{\rm NN}} = 200$~GeV, computed using the \texttt{AMPT-SM} model. The calculations employ a pseudorapidity gap $\eta_{\rm gap} = 0.4$ within $|\eta| \leq 1.0$ to suppress short-range non-flow correlations. At low-$p_{\rm T}$ ($\lesssim 0.5$~GeV/$c$), $v_{0}(p_{\rm T})$ is negative across all centrality intervals, consistent with the anti-correlation between event-by-event (EbyE) mean-$p_{\rm T}$ fluctuations and the differential particle yield at a given $p_{\rm T}$~\cite{schenke2020,parida2024}. This behavior reflects that events with larger-than-average mean transverse momentum 
($[p_{\rm T}] > \langle [p_{\rm T}] \rangle$), corresponding to stronger radial expansion, produce flatter $p_{\rm T}$ spectra with enhanced yields at high-$p_{\rm T}$ and suppressed yields at low-$p_{\rm T}$, while events with weaker radial expansion exhibit the opposite trend~\cite{parida2024}. For $p_{\rm T} > 0.55$~GeV/$c$, $v_{0}(p_{\rm T})$ turns positive and rises non-linearly with $p_{\rm T}$. The magnitude of $v_{0}(p_{\rm T})$ is significantly larger for the 70--80\% centrality class than for 0--5\% and 30--40\%, consistent with the expectation that \textit{peripheral} collisions exhibit larger EbyE geometric fluctuations, which drive stronger radial flow fluctuations~\cite{bozek2012, aad2024xexeatlas}. Crucially, the three $p_{\rm T}^{\rm ref}$ ranges yield mutually consistent results for $p_{\rm T} \lesssim 1.6$~GeV/$c$, demonstrating that $v_{0}(p_{\rm T})$ factorizes from the choice of reference range in this regime (a direct signature of collective radial flow) while deviations emerging at higher-$p_{\rm T}$ indicate the onset of non-flow contributions~\cite{aad2026atlas,acharya2026alice}. One can see that $v_{0}(p_{\rm T})$ magnitude for $p_{\rm T}^{\rm ref}: 0.2-1.5$ GeV dips at higher-$p_{\rm T}$, indicating that it is least affected by the non-flow correlations.
\subsection{Centrality dependence of $v_{0}(p_{\rm T})$}

Fig.~\ref{fig:v0(pT)_v0(pT)/v0_v0(pT)sqrt{Nch}} ($\textit{left}$) presents the $p_{\rm T}$ dependence of $v_{0}(p_{\rm T})$ for inclusive charged hadrons in Au+Au collisions at $\sqrt{s_{\rm NN}} = 200$ GeV, obtained from \texttt{AMPT-SM} simulations across nine centrality classes viz. 0-5\%, 5-10\%, 10-20\%, 20-30\%, 30-40\%, 40-50\%, 50-60\%, 60-70\%, and 70-80\%. 
A clear centrality dependence is observed over the entire $p_{\rm T}$ range. The \textit{most central} events (0--5\%) exhibit the smallest values of $v_{0}(p_{\rm T})$, whereas the \textit{most peripheral} class (70--80\%) shows the largest magnitude, particularly at higher-$p_{\rm T}$. The increase of $v_{0}(p_{\rm T})$ with $p_{\rm T}$ is modest for \textit{central} collisions, whereas a more pronounced rise is observed in \textit{peripheral} events. This systematic evolution can be understood in terms of the interplay between collision geometry and collective expansion dynamics. Although \textit{central} collisions generate larger and longer-lived systems with stronger average radial flow, the relative magnitude of flow-induced fluctuations, as quantified by $v_{0}(p_{\rm T})$, is reduced due to the larger participant multiplicity, which effectively averages out event-by-event fluctuations.

Fig.~\ref{fig:v0(pT)_v0(pT)/v0_v0(pT)sqrt{Nch}} (\textit{center}) shows the $p_{\rm T}$ dependence of the scaled radial flow observable $v_{0}(p_{\rm T})/v_{0}$ for different centrality classes, where $v_{0}$ denotes the corresponding $p_{\rm T}$-integrated radial flow magnitude (See Eq.~(\ref{eq:5})). The normalized distributions $v_{0}(p_{\rm T})/v_{0}$ exhibit a clear centrality-independent scaling behavior up to $p_{\rm T}\lesssim 2$~GeV/$c$. 
This scaling suggests that, in this momentum region, the correlations are predominantly governed by bulk collective dynamics within the \texttt{AMPT-SM} framework. A qualitatively similar scaling trend has been reported by the ATLAS Collaboration in Pb+Pb collisions at $\sqrt{s_{\rm NN}} = 5.02$~TeV~\cite{aad2026atlas}, indicating that this feature persists across collision energies. At higher transverse momentum, $p_{\rm T}\gtrsim 2$~GeV/$c$, systematic deviations from the low-$p_{\rm T}$ scaling trend become apparent, particularly in \textit{peripheral} collisions where $v_{0}(p_{\rm T})/v_{0}$ continues to increase rather than following the universal behavior observed at lower-$p_{\rm T}$. These deviations are consistent with a reduced dominance of collective expansion and an increasing relative contribution from non-collective particle production mechanisms, such as hard partonic scatterings and jet fragmentation.

To further probe the origin of radial flow fluctuations, we examine 
$v_{0}(p_{\rm T})\sqrt{\rm N_{\rm ch}}$ as a function of $p_{\rm T}$ for different centralities in Fig.~\ref{fig:v0(pT)_v0(pT)/v0_v0(pT)sqrt{Nch}} (\textit{right}). An approximate collapse of the curves is observed at low-$p_{\rm T}$ (up to $\sim 1.2$~GeV/$c$), consistent with a $1/\sqrt{\rm N_{\rm ch}}$ scaling in this momentum region. A similar trend has been reported in Pb+Pb collisions at $\sqrt{s_{\rm NN}} = 5.02$~TeV and interpreted within an effective independent source framework~\cite{aad2026atlas}. In the present \texttt{AMPT-SM} calculations, which incorporate partonic scatterings followed by quark coalescence and hadronic rescattering, the same scaling behavior persists despite the presence of strong interactions in the partonic phase. The persistence of the scaling in this interacting system suggests that EbyE fluctuations of the collective radial expansion are progressively diluted with increasing multiplicity. At higher-$p_{\rm T}$, the scaling gradually breaks down, reflecting the increasing contribution of harder partonic processes that are less governed by bulk collective dynamics.

\begin{figure*}
    \centering
    \includegraphics[width=0.9\linewidth]{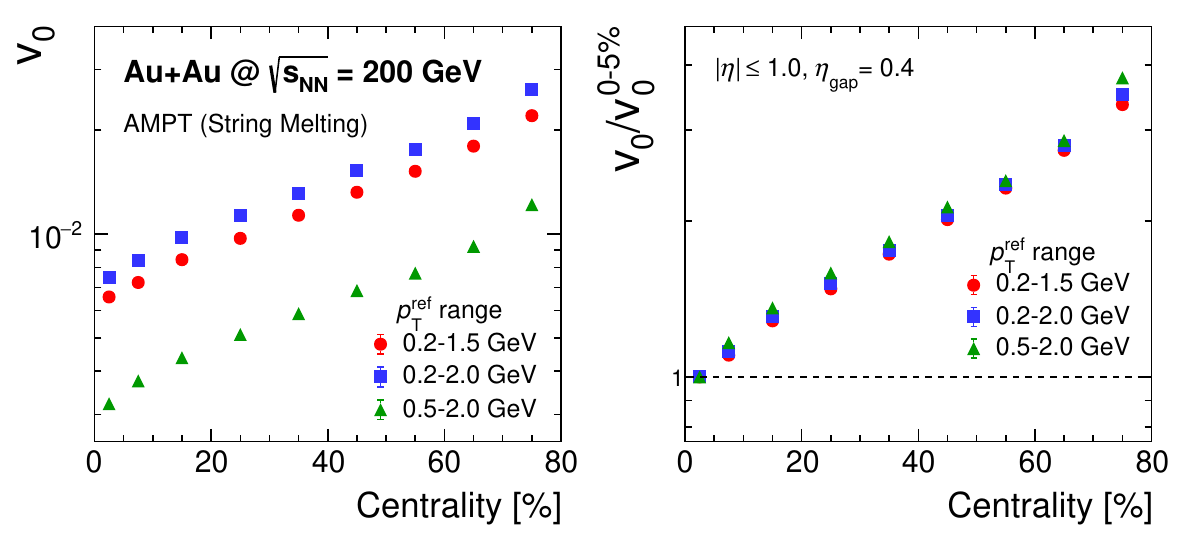}
    \includegraphics[width=0.9\linewidth]{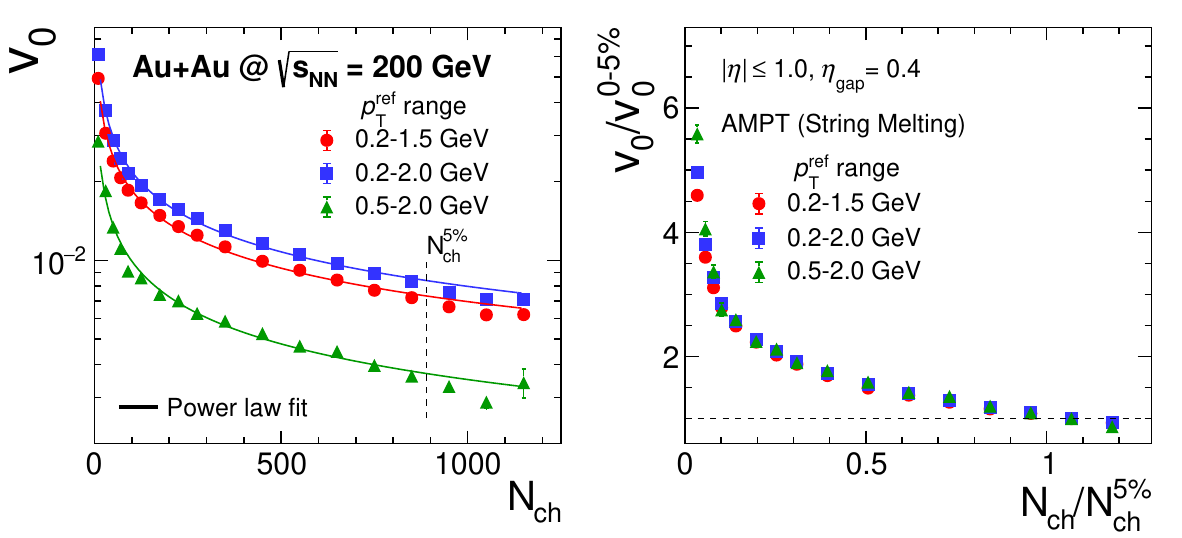}
    \caption{(Color online) Centrality dependence of integrated radial flow measure, $v_{0}$ (\textit{left panel}) and $v_{0}/v_{0}^{0-5\%}$ (\textit{right panel}) for inclusive charged hadrons at three different $p_{\rm T}^{\rm ref}$ ranges for Au+Au collisions at $\sqrt{s_{\rm NN}} = 200$ GeV simulated using \texttt{AMPT-SM} framework.}
    \label{fig:v0}
\end{figure*}

\begin{figure*}
    \centering
    \includegraphics[width=\linewidth]{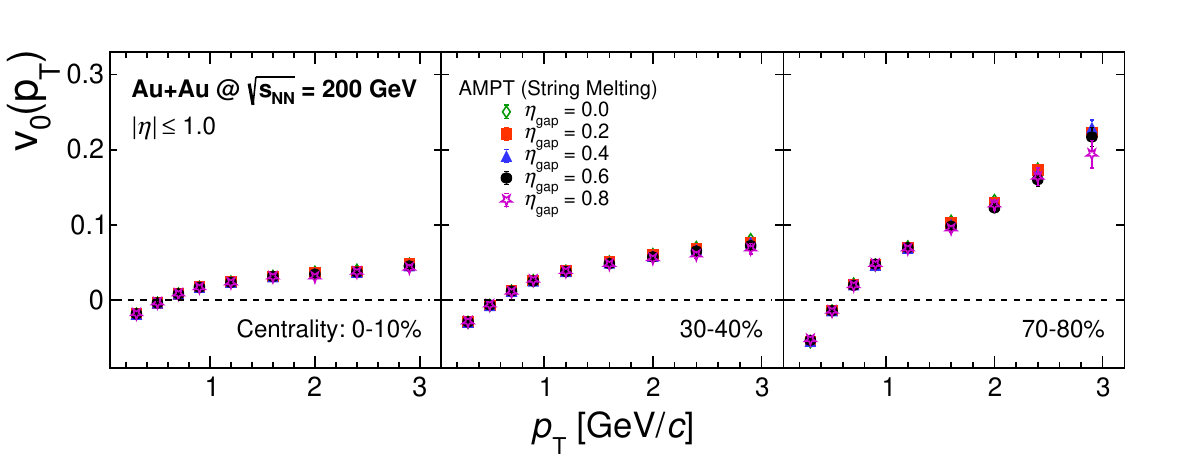}
    \caption{(Color online) Pseudorapidity gap dependence of inclusive $v_{0}(p_{\rm T})$ within three centrality intervals: 0-10\% (\textit{most central}), 30-40\% (\textit{semi-central}) and 70-80\% (\textit{peripheral}) for Au+Au collisions at $\sqrt{s_{\rm NN}} = 200$ GeV simulated using the \texttt{AMPT-SM} framework.}
    \label{fig:pseudorapidity_gap}
\end{figure*}

\begin{figure*}
    \centering
    \includegraphics[width=\linewidth]{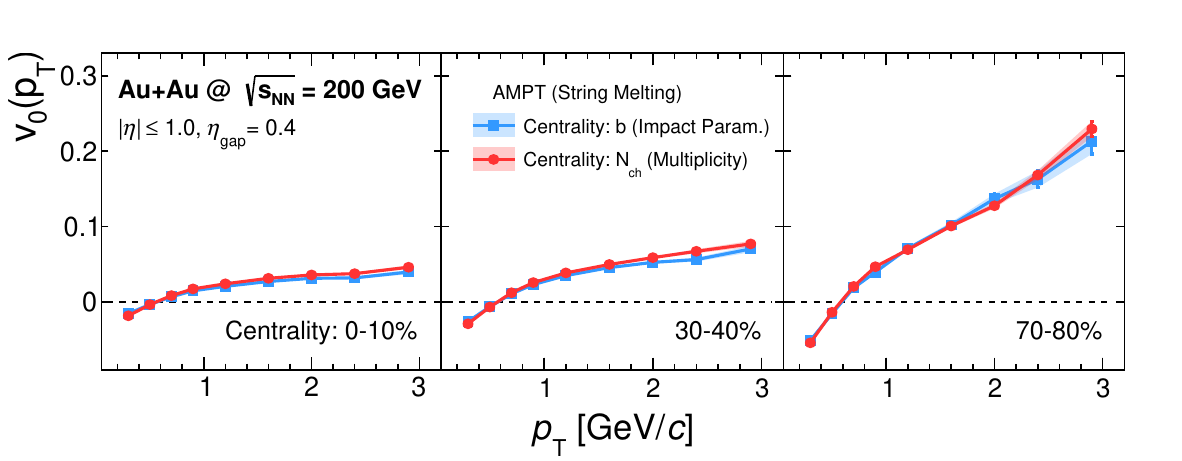}
    \caption{(Color online) Dependence of $v_{0}(p_{\rm T})$ on the choice of centrality estimators for \textit{most central}, \textit{mid-central} and \textit{peripheral} Au+Au collisions generated using \texttt{AMPT-SM} at $\sqrt{s_{\rm NN}} = 200$ GeV.}
    \label{fig:effect_cent_est}
\end{figure*}
\subsection{Centrality dependence of integrated radial flow fluctuations}

The \textit{top left panel} of Fig.~\ref{fig:v0} presents the centrality evolution of the integrated radial flow observable $v_{0}$ (calculated using Eq.~(\ref{eq:5})) in Au+Au collisions at $\sqrt{s_{\rm NN}} = 200$ GeV for three reference transverse momentum ranges: $p_{\rm T}^{\rm ref} = 0.2-1.5~\rm GeV, 0.2-2.0~\rm GeV$ and $0.5-2.0~\rm GeV$. The results reveal a systematic increase in $v_{0}$ from central to peripheral collisions across all $p_{\rm T}^{\rm ref}$ selections, consistent with the expected behavior of event-by-event (EbE) radial flow fluctuations in heavy-ion collisions~\cite{aad2026atlas,bhatta2025disentangling}. The magnitude of $v_{0}$ exhibits a clear dependence on the chosen reference momentum window, with the widest range ($0.2-2.0~\rm GeV$) yielding the largest values. This behavior follows directly from the sum rule relating $v_{0}$ to $v_{0}(p_{\rm T})$, where broader integration ranges encompass larger portions of the momentum spectrum affected by radial flow fluctuations~\cite{aad2026atlas}. The range ($0.5-2.0~ \rm GeV$) produces correspondingly smaller $v_{0}$ values, reflecting the reduced contribution from low-momentum particles where radial flow effects are most pronounced.

The \textit{top right panel} of Fig.~\ref{fig:v0} shows $v_{0}$ normalized to its value in the 0–5\% most central collisions, $v_{0}/v_{0}^{0-5\%}$. The three $p_{\rm T}^{\rm ref}$ selections exhibit similar normalized trends, with the corresponding curves overlapping within systematic uncertainties. This approximate scaling indicates that although the absolute magnitude of radial flow fluctuations depends on the chosen momentum interval, the centrality dependence follows a common underlying pattern that is largely independent of the specific $p_{\rm T}^{\rm ref}$ selection. Within the string melting configuration of \texttt{AMPT}, this scaling can be attributed to event-by-event fluctuations in the effective transverse expansion velocity generated during the partonic cascade stage. Fluctuations in the initial parton density and subsequent scattering history coherently modify the transverse momentum spectra during the partonic cascade stage. Hence, $v_{0}$ captures event-by-event fluctuations of the global spectral shape, consistent with a collective origin of radial flow fluctuations generated in the partonic evolution. The present findings are consistent with the results reported for Pb+Pb collisions at $\sqrt{s_{\rm NN}} = 5.02$ TeV at the LHC by the ATLAS Collaboration~\cite{aad2026atlas}, where a similar scaling behavior of $v_{0}$ was observed, supporting its interpretation as a measure of collective radial flow fluctuations.

The same analysis is presented in the \textit{bottom panels} of Fig.~\ref{fig:v0}, where $\rm N_{ch}$ is used as the centrality variable. Since $\rm N_{ch}$ is inversely correlated with centrality, $v_{0}$ exhibits a monotonic decrease with increasing multiplicity, consistent with the trends observed in the centrality-based representation. The dependence on the chosen $p_{\rm T}^{\rm ref}$ interval remains unchanged, with the widest momentum range yielding the largest magnitude of $v_{0}$. The \textit{bottom right panel} shows $v_{0}$ normalized to its value at $\rm N_{ch}^{5\%}$ as a function of $\rm N_{ch}/N_{ch}^{5\%}$. The three $p_{\rm T}^{\rm ref}$ selections again display approximate scaling within uncertainties, demonstrating that the multiplicity-based representation preserves the same underlying fluctuation pattern. Quantitatively, the multiplicity dependence is well described by a power-law parametrization, $v_{0}=AN_{\rm ch}^{b}$, for all $p_{\rm T}^{\rm ref}$ intervals. The extracted exponents, summarized in Tab.~\ref{tab:v0_fit_parameters}, lie in the range $-0.42$ to $-0.45$ with only weak momentum dependence. For comparison, the corresponding measurements reported by the ATLAS Collaboration were fitted with the same functional form (See Fig.~\ref{fig:ATLAS}), yielding exponents close to $b \approx -0.40$ to $-0.42$. In both cases, the magnitude of $b$ is consistent with an approximate inverse square-root dependence on $\rm N_{\rm ch}$, indicating that relative radial flow fluctuations decrease with increasing multiplicity. The overall agreement in scaling behavior between model and data supports a common multiplicity-driven origin of radial flow fluctuations, consistent with the collective nature of the underlying transverse expansion.

\begin{table}[h]
\centering
\caption{Power-law fit parameters for the $v_0$ dependence in \texttt{AMPT-SM} Au+Au collisions (\textit{top}) and ATLAS Pb+Pb data~\cite{aad2026atlas} (\textit{bottom}), obtained using $v_0 = A\,N_{\rm ch}^{\,b}$.}
\label{tab:v0_fit_parameters}
\begin{tabular}{lccc}
\toprule\toprule
System ($\sqrt{s_{\rm NN}}$) & $p_{\rm T}^{\rm ref}$ [GeV] & $A$ & $b$ \\
\midrule
& $0.2$--$1.5$ & $0.130 \pm 0.001$ & $-0.424 \pm 0.002$ \\
Au+Au ($200$ GeV) & $0.2$--$2.0$ & $0.164 \pm 0.001$ & $-0.438 \pm 0.001$ \\
& $0.5$--$2.0$ & $0.080 \pm 0.002$ & $-0.452 \pm 0.005$ \\
\midrule
& $0.5$--$2.0$ & $0.185 \pm 0.011$ & $-0.421 \pm 0.008$ \\
Pb+Pb ($5.02$ TeV) & $0.5$--$5.0$ & $0.316 \pm 0.017$ & $-0.411 \pm 0.007$ \\
& $1.0$--$5.0$ & $0.204 \pm 0.011$ & $-0.408 \pm 0.007$ \\
\bottomrule\bottomrule
\end{tabular}
\end{table}

\subsection{$\eta_{\rm gap}$ dependence of $v_{0}(p_{\rm T})$ }

To isolate the contributions from short-range correlations, we investigate the pseudorapidity gap dependence of $v_{0}(p_{\rm T})$ in Fig.~\ref{fig:pseudorapidity_gap} for three centrality classes (0--10\%, 30--40\%, and 70--80\%). The observable exhibits the characteristic sign change, being negative at low-$p_{\rm T}$ ($< 0.5$~GeV/$c$) and positive at higher-$p_{\rm T}$ ($> 0.5$~GeV/$c$). The weak dependence on $\eta_{\rm gap}$, particularly in \textit{central} and \textit{mid-central} collisions, indicates that $v_{0}(p_{\rm T})$ is predominantly driven by long-range collective correlations rather than short-range nonflow effects. In \textit{peripheral} collisions (70--80\%), a mild $\eta_{\rm gap}$ dependence emerges at higher-$p_{\rm T}$, suggesting a modest contribution from jet-like correlations where the system size and collective effects are reduced. These observations support the interpretation that radial flow in Au+Au collisions at RHIC energies exhibits collective behavior consistent with recent measurements in Pb+Pb collisions at the LHC~\cite{acharya2026alice,aad2026atlas}.\\

\subsection{Effect of centrality estimator on $v_{0}(p_{\rm T})$}

In this section, we compare two commonly used centrality estimators: midrapidity charged-particle multiplicity ($|\eta| \leq 0.5$) and the impact parameter $b$. The impact parameter directly characterizes the  geometric overlap of the colliding nuclei, whereas the multiplicity-based 
estimator mimics the experimental procedure in which centrality is determined from particle yields measured within a given detector acceptance. Fig.~\ref{fig:effect_cent_est} shows $v_{0}(p_{\rm T})$ as a function of $p_{\rm T}$ for three representative centrality classes (0--10\%, 30--40\%, and 70--80\%) obtained using both estimators. We observe that at low-$p_{\rm T}$, $v_{0}(p_{\rm T})$ is nearly identical for both centrality estimators across all centrality classes. At higher-$p_{\rm T}$, small but systematic differences emerge, with the impact parameter based selection yielding slightly smaller values of $v_{0}(p_{\rm T})$. This behavior can also be understood in terms of residual volume fluctuations associated with multiplicity-based centrality selection. Within a given multiplicity interval, event-by-event variations in entropy deposition lead to fluctuations in the effective system size, which couple to variations of the spectral shape. Such volume-induced fluctuations can modestly enhance the extracted $v_{0}(p_{\rm T})$, particularly at high-$p_{\rm T}$ where relative 
spectral modifications induced by variations in collective transverse expansion become more pronounced. In contrast, selecting events by the impact parameter constrains the collision geometry more directly and reduces the influence of residual volume fluctuations.
\begin{figure*}
    \centering
    \includegraphics[width=\linewidth]{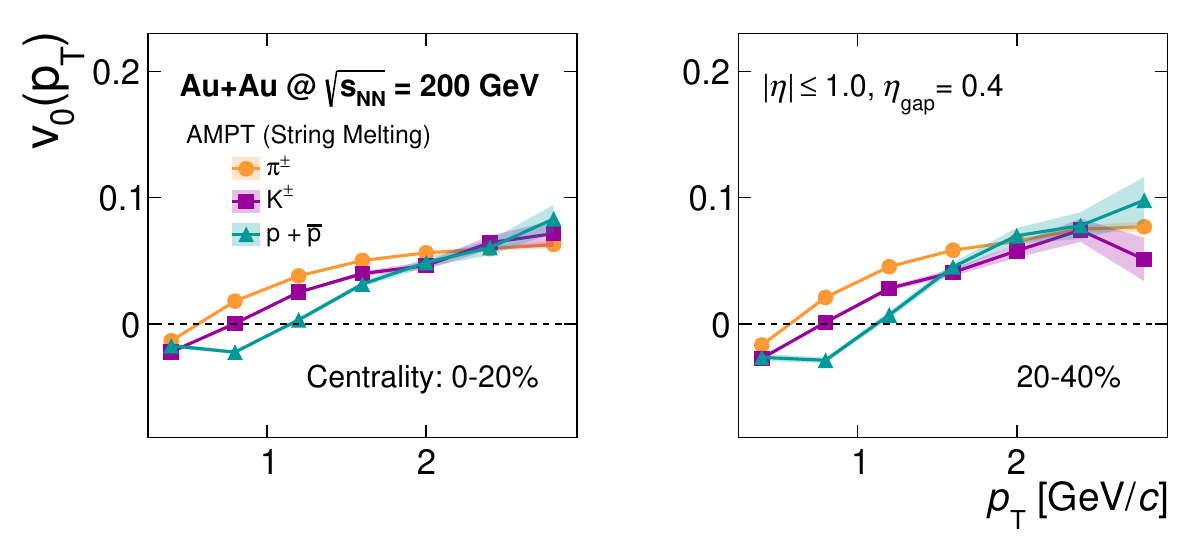}
        \caption{(Color online) Transverse momentum ($p_{\rm T}$) dependence of $v_{0}(p_{\rm T})$ for identified hadrons in Au+Au collisions generated with the String Melting version of \texttt{AMPT} model, shown for 0-20\% (\textit{most central}) and 20-40\% (\textit{semi-central}) centrality selections. \textit{Shaded bands} indicate bootstrap-estimated statistical uncertainties.}
    \label{fig:auau_nonscaled}
\end{figure*}

\begin{figure*}
    \centering
    \includegraphics[width=\linewidth]{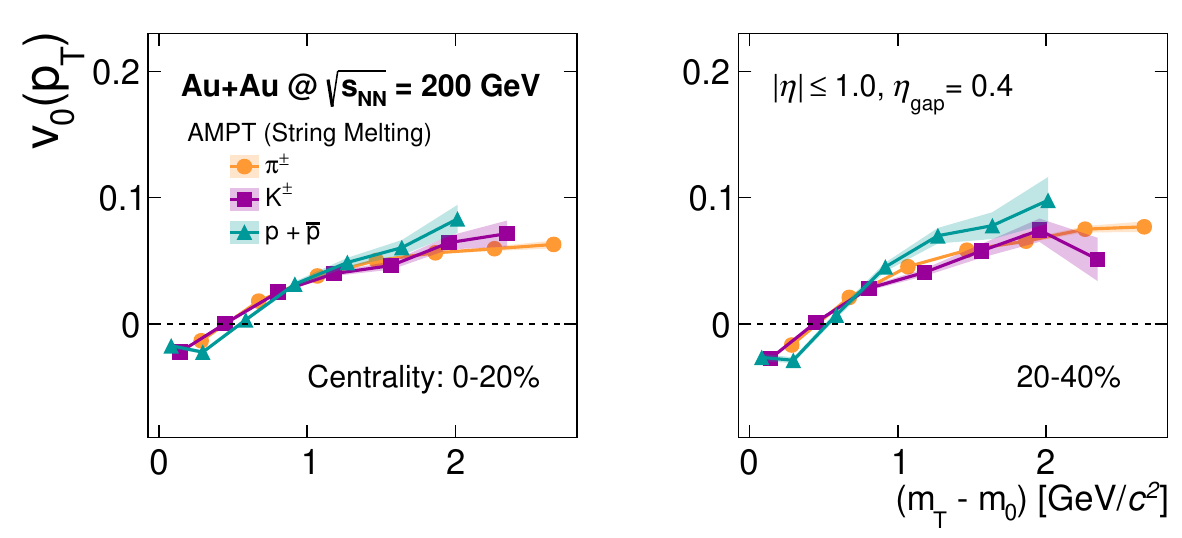}
        \caption{ (Color online) The $v_{0}(p_{\rm T})$ observable as a function of transverse kinetic energy $\rm (m_{T} - m_{0})$ for identified hadrons in \texttt{AMPT-SM} simulations of Au+Au collisions, displayed for 0-20\% (\textit{most central}) and 20-40\% (\textit{semi-central}) centrality bins. Bootstrap-derived statistical uncertainties are shown as \textit{shaded regions}.}
    \label{fig:auau_mTminusm0_nonscaled}
\end{figure*}

\begin{figure*}
    \centering
    \includegraphics[width=\linewidth]{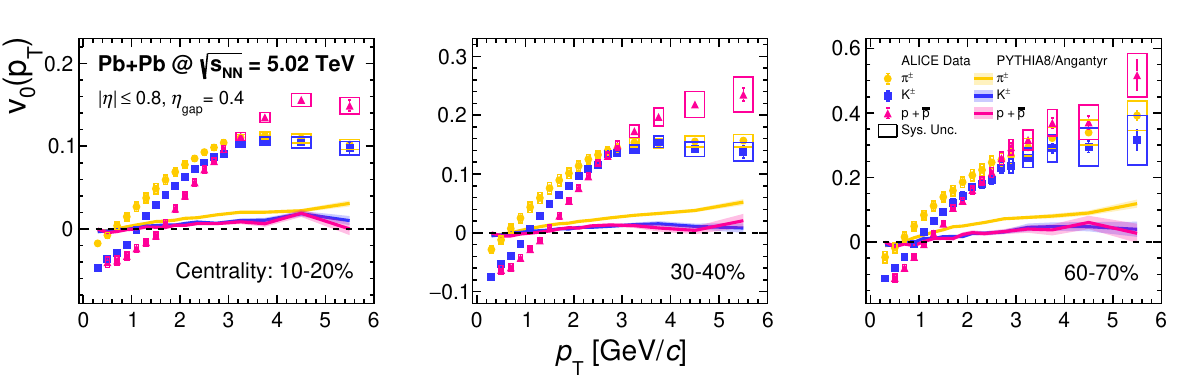}
    \caption{(Color online) Comparison of $v_{0}(p_{\rm T})$ between \texttt{PYTHIA8/Angantyr} predictions and ALICE measurements~\cite{acharya2026alice} for identified hadrons in Pb+Pb collisions at $\sqrt{s_{\rm NN}}$ = 5.02 TeV, presented for 10-20\%, 30-40\% and 60-70\% centrality intervals. Model calculations are represented by \textit{solid curves} with \textit{shaded bands} denoting bootstrap uncertainties, while experimental data are shown as \textit{filled markers} with uncertainties shown in \textit{vertical bars} (statistical) and \textit{hollow boxes} (systematic).}
    \label{fig:angantyr_results}
\end{figure*}

\begin{figure*}
    \centering
    \includegraphics[width=\linewidth]{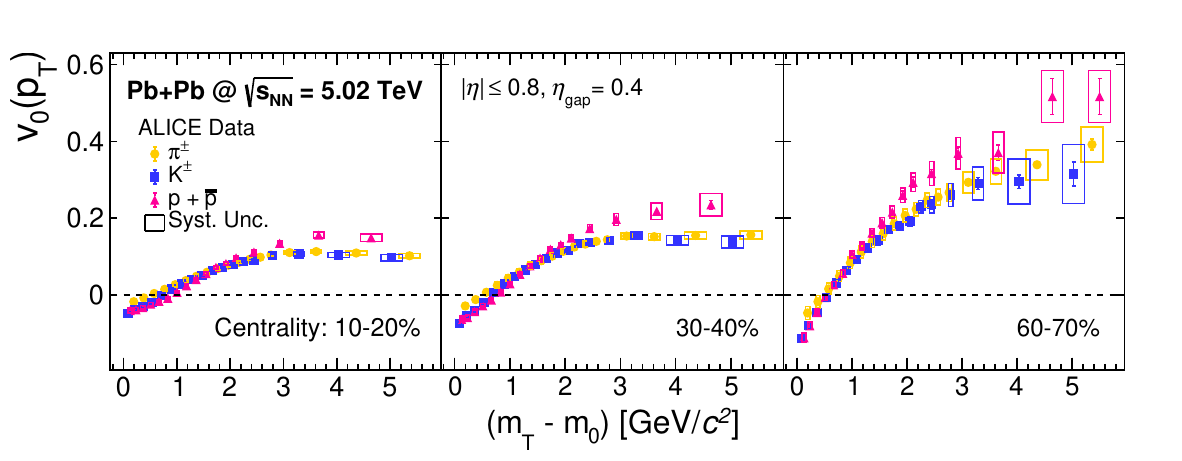}
    \caption{(Color online) $v_{0}(p_{\rm T})$ of identified hadrons presented in terms of transverse kinetic energy, ($m_{\rm T} - m_{\rm 0}$) for Pb+Pb collisions at $\sqrt{s_{\rm NN}} = 5.02$ TeV. The plot is compiled from the experimental data of ALICE~\cite{acharya2026alice}.}
    \label{fig:v0pTversusmTminusm0_alice}
\end{figure*}

\subsection{$v_{0}(p_{\rm T})$ for identified charged hadrons at $\sqrt{s_{\rm NN}} = 200$ $\rm GeV$} \label{sec:4.1}

In Fig.~\ref{fig:auau_nonscaled}, we display the radial flow observable, $v_{0}(p_{\rm T})$ for identified hadrons ($\pi^{\pm}$, $K^{\pm}$, and $p+\overline{p}$) in Au+Au collisions at $\sqrt{s_{\rm NN}} = 200$~GeV, obtained from \texttt{AMPT-SM} simulations for \textit{central} (0-20\%) and \textit{mid-central} (20-40\%) class. The observable is calculated using the subevent method with $\eta_{\rm gap}$ = 0.4 to minimize the short-range non-flow correlations. A clear mass ordering is observed at lower-$p_{\rm T}$ ($p_{\rm T} \leq 2$ GeV/\textit{c}), while a characteristic baryon–meson crossover emerges at intermediate $p_{\rm T}$ ($2\leq p_{\rm T} \leq 3$ GeV/\textit{c}). These features are consistent with expectations from collective radial expansion of the produced medium. The trends align with measurements at $\sqrt{s_{\rm NN}} = 5.02$ TeV by ALICE~\cite{acharya2026alice} and with hydrodynamic model calculations~\cite{du2026}.

To further probe the collective nature of particle emission, the dependence of $v_{0}(p_{\rm T})$ on transverse kinetic energy $(m_{\rm T}-m_{\rm 0})$ is studied, which provides a stringent test of the mass-ordering observed at low-$p_{\rm T}$. Fig.~\ref{fig:auau_mTminusm0_nonscaled} shows $v_{0}(p_{\rm T})$ plotted against $(m_{\rm T}-m_{0})$ for the 0-20\% and 20-40\% centrality classes in Au+Au collisions at $\sqrt{s_{\rm NN}}= 200$ GeV. An approximate mass scaling is observed at lower-$(m_{\rm T}-m_{0})$ (up to about 1 GeV/$c^{2}$), while a clear meson-baryon separation emerges at higher $(m_{\rm T}-m_{\rm 0})$. This behavior is consistent with hydrodynamic-like pressure gradients developed in the early partonic phase, as captured by the \texttt{AMPT-SM} framework.
\subsection{Estimation of $v_{0}(p_{\rm T})$ at $\sqrt{s_{\rm NN}} = 5.02$ TeV}

This analysis is further extended to Pb+Pb collisions at $\sqrt{s_{\rm NN}}$ = 5.02 TeV by comparing the radial flow observable $v_{0}(p_{\rm T})$ for identified charged hadrons from \texttt{PYTHIA8/Angantyr} simulations with the corresponding ALICE data~\cite{acharya2026alice} for 10–20\% (\textit{central}), 30–40\% (\textit{semi-central}) and 60-70\% (\textit{peripheral}) classes, as shown in Fig.~\ref{fig:angantyr_results}. As reported by the ALICE collaboration~\cite{acharya2026alice}, the experimental results display clear signatures of collective expansion, characterized by a pronounced mass ordering at low-$p_{\rm T}$ ($p_{\rm T} \leq 3~\text{GeV}/\textit{c}$) and a distinct meson-baryon crossing at intermediate-$p_{\rm T}$ ($p_{\rm T} \geq 3~\text{GeV}/\textit{c}$). In contrast, the $v_{0}(p_{\rm T})$ values predicted by \texttt{PYTHIA8/Angantyr} remain close to zero over the entire $p_{\rm T}$ range, exhibiting neither mass ordering nor meson-baryon crossing. This pronounced discrepancy reflects the absence of a hydrodynamic evolution stage in the model, which prevents the buildup of sustained pressure gradients required to generate collective radial flow.  The inability of the model to reproduce the observed experimental features provides further evidence that the measured $v_{0}(p_{\rm T})$ patterns arise from genuine final-state collective dynamics, rather than from initial-state effects or string-fragmentation mechanisms.

Fig.~\ref{fig:v0pTversusmTminusm0_alice} examines the measured $v_{0}(p_{\rm T})$ data~\cite{acharya2026alice}, replotted in terms of the transverse kinetic energy $(m_{\rm T}-m_{0})$ computed from $p_{\rm T}$, for Pb+Pb collisions at $\sqrt{s_{\rm NN}}=5.02$ TeV. An approximate mass scaling is observed at low-$(m_{\rm T}-m_{0})$, up to $\sim$2.5, 2.0, and 1.5 GeV/$c^{2}$ for \textit{central}, \textit{semi-central}, and \textit{peripheral} collisions, respectively, followed by a clear separation between meson and baryon branches at higher transverse kinetic energies. The observed systematics closely resemble those obtained in the \texttt{AMPT-SM} calculations presented above (Fig.~\ref{fig:auau_mTminusm0_nonscaled}), and are consistent with established elliptic-flow ($v_{2}$) measurements at RHIC energies reported by the STAR and PHENIX Collaborations~\cite{adams2004,adams2005multistrange,adare2007phenix}. While these features confirm the collective origin of the observed radial flow, they alone cannot distinguish whether the collectivity develops predominantly in the partonic or hadronic stage. To address this question, we perform, for the first time, a \textit{Number of Constituent Quark} (NCQ) scaling analysis of the $v_{0}(p_{\rm T})$ observable at RHIC energy for Au+Au collisions, as detailed in the following section.
\begin{figure*}
    \centering
        \includegraphics[width=\linewidth]{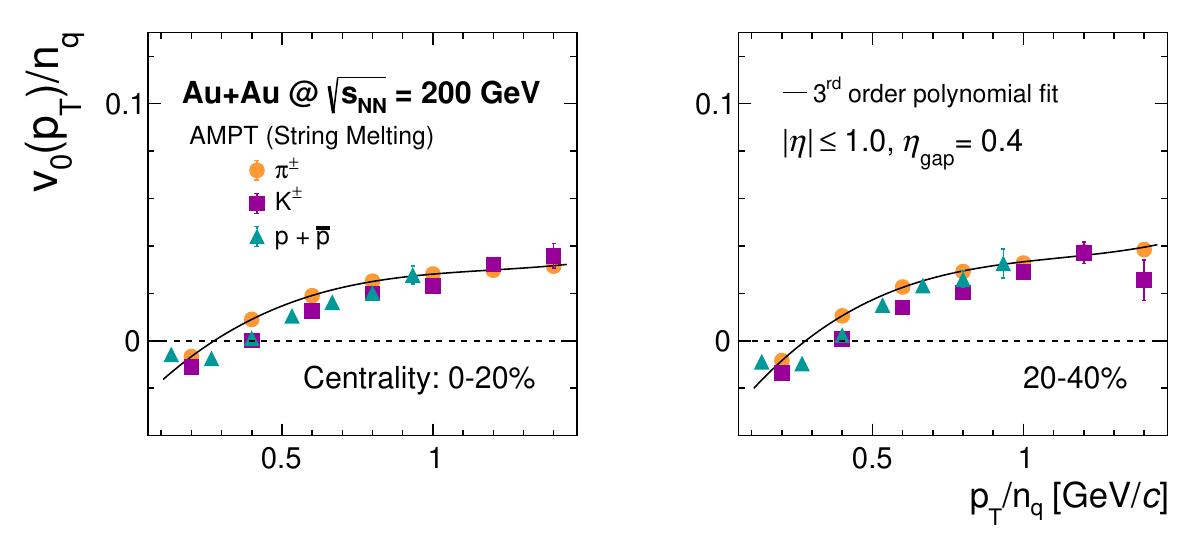}
        \includegraphics[width=\linewidth]{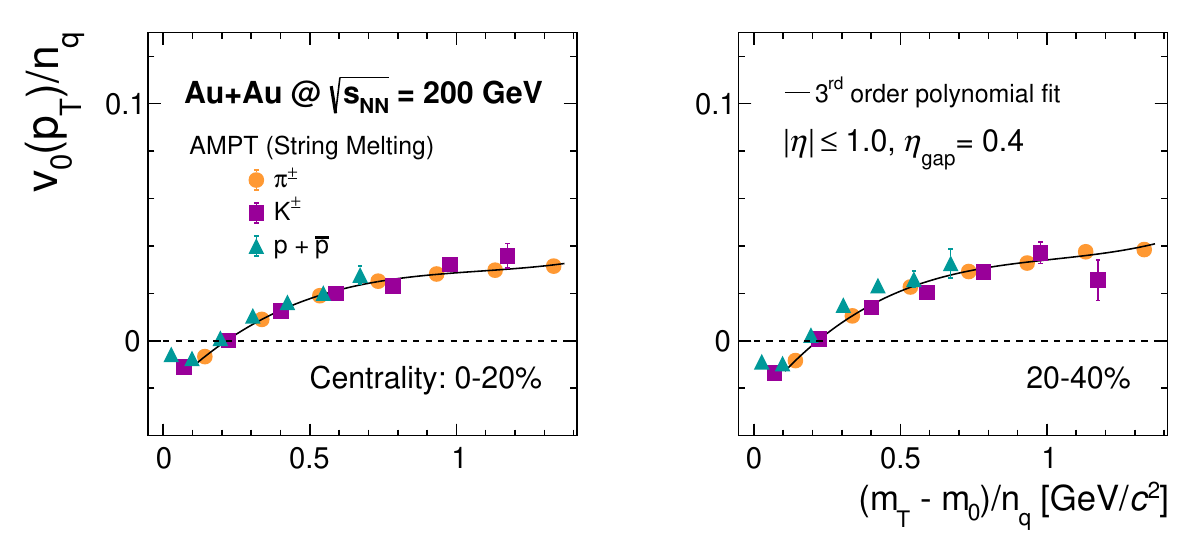}
        \caption{(Color online) \textit{Number of Constituent Quark} (NCQ) scaled radial flow observable, $v_{0}(p_{\rm T})/n_{q}$, versus $p_{\rm T}/n_{q}$ (\textit{top panel}) and $(m_{T} - m_{0})/n_{q}$ (\textit{bottom panel}) for identified hadrons in \texttt{AMPT-SM} simulations of Au+Au collisions at 0-20\% (\textit{most central}) and 20-40\% (\textit{semi-central}) centrality. A third-order polynomial fit to $\pi^{\pm}$ data (\textit{solid curve}) serves as the NCQ scaling reference. Bootstrap uncertainties are displayed as vertical error bars.}
    \label{fig:auau_scaled}
\end{figure*}

\begin{figure*}
    \includegraphics[width=\linewidth]{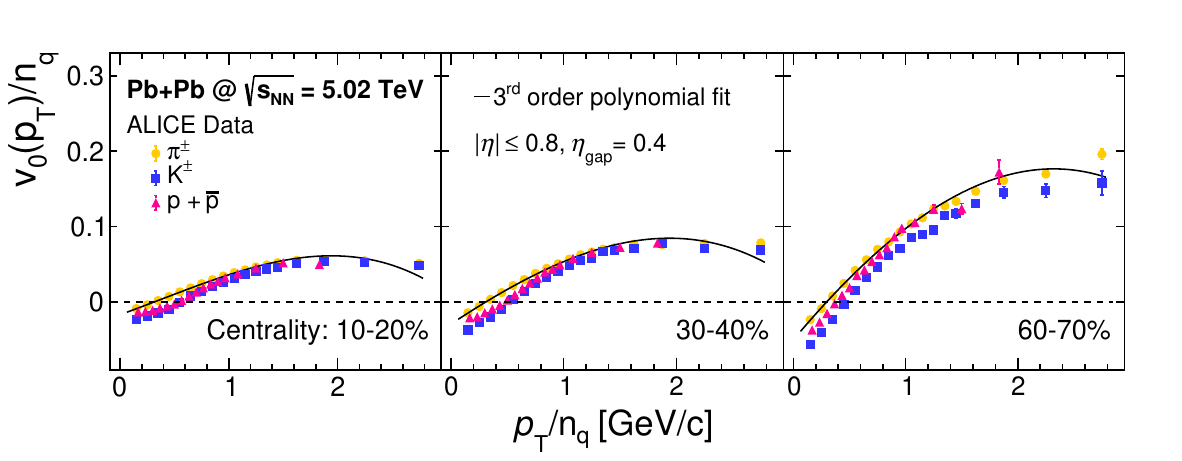}
    \includegraphics[width=\linewidth]{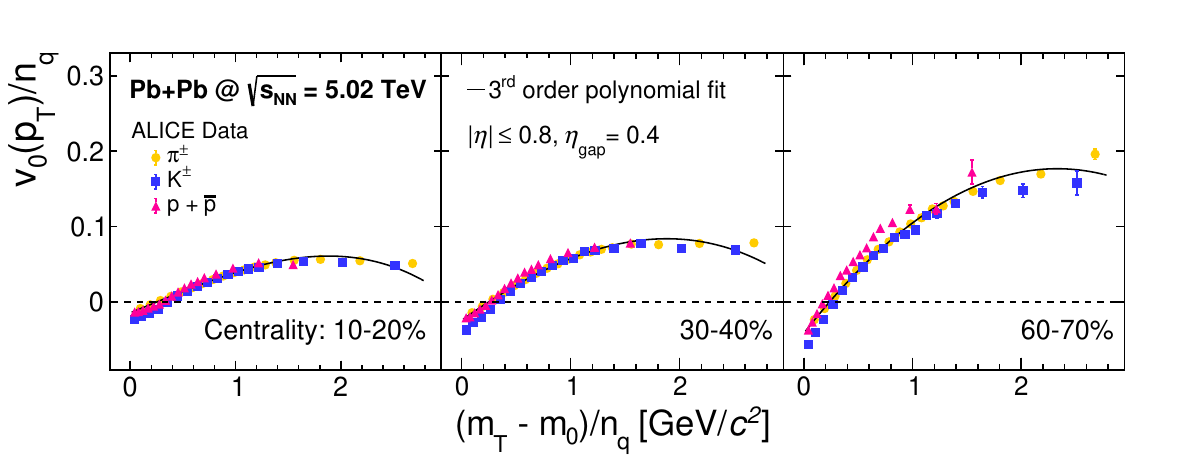}
     \caption{(Color online) NCQ-scaled $v_{0}(p_{\rm T})$ of identified hadrons as a function of NCQ-scaled $p_{\rm T}$ (\textit{top panel}) and as a function of NCQ-scaled ($ m_{\rm T} - m_{\rm 0}$) (\textit{bottom panel}) in Pb+Pb collisions at $\sqrt{s_{\rm NN}} = 5.02$ TeV. These plots are compiled using the experimental data of ALICE~\cite{acharya2026alice}. \textit{Vertical bars} represent statistical uncertainties.}
    \label{fig:NCQ_scaled_alice_data}
\end{figure*}

\subsection{NCQ scaling of radial flow observable $v_{0}(p_{\rm T})$} \label{sub_sec:ncq_scaling}

\subsubsection{NCQ scaling at $\sqrt{s_{\rm NN}} = 200$ $\rm GeV$} \label{subsub_sec:ncq_scaling_AuAu}

We first apply \textit{Number of Constituent Quark} (NCQ) scaling to the $v_{0}(p_{\rm T})$ observable by rescaling both the flow magnitude and the kinematic variables with the number of constituent quarks $n_q$. Fig.~\ref{fig:auau_scaled} presents the resulting scaled distributions, $v_{0}(p_{\rm T})/n_q$, for Au+Au collisions at $\sqrt{s_{\rm NN}} = 200$ GeV simulated using the \texttt{AMPT-SM} model. The results are shown as functions of $p_{\rm T}/n_q$ (\textit{top}) and $(m_{\rm T}-m_{0})/n_q$ (\textit{bottom}) for \textit{central} (0--20\%) and \textit{semi-central} (20--40\%) events. While the scaling with $p_{\rm T}/n_q$ is only approximate, a visibly improved scaling is observed when the transverse kinetic energy variable $(m_{\rm T}-m_{\rm 0})/n_q$ is employed. Moreover, the degree of NCQ scaling is found to be systematically better for \textit{central} (0--20\%) collisions compared to the \textit{semi-central} (20--40\%) class, indicating a stronger manifestation of collectivity in more \textit{central} events.

To quantify the observed scaling behavior, we have adopted the following procedure. The $v_{0}(p_{T})$ spectrum of pions ($\pi^{\pm}$) is fitted with a third-order polynomial (solid line), which is then used as a reference to extract a scaling parameter $S_f$. This parameter measures the deviation of the kaon ($K^{\pm}$) and proton ($p+\bar{p}$) $v_{0}(p_{T})$ distributions from the pion reference. $S_f$ is therefore defined as,
\begin{equation} \label{eq:scaling_factor}
    S_{f} = \sum_{i = 1}^{2} \sum_{j = 1}^{N} |y_{j} - f(x_{j})|_{i}
\end{equation}
where $N$ is the number of bins in which $v_{0}(p_{\rm T})$ is calculated for $K^{\pm}$ and $p+\overline p$ and $y_{j}$ correspond to $v_{0}(p_{\rm T})$ values of $K^{\pm}$ and $p + \overline p$. By definition, for ideal scaling, $S_{f}$ must be equal to zero.

The extracted $S_f$ values for NCQ scaling with respect to $p_{\rm T}/n_q$ and $(m_{\rm T}-m_{0})/n_q$ for different centrality classes are summarized in Tab.~\ref{tab:scaling}. While the qualitative scaling features have already been discussed above, the $S_f$ values provide a quantitative validation of these observations. Two clear trends emerge: (i) the scaling improves significantly when $(m_{\rm T}-m_{0})/n_q$ is used as the scaling variable, and (ii) the scaling is systematically better for more \textit{central} collisions.

\begin{table}[h]
\centering
\caption{Scaling parameter $S_f$ for scaled \texttt{AMPT} Au+Au results (\textit{top}) and scaled ALICE~\cite{acharya2026alice} Pb+Pb results (\textit{bottom}).}
\label{tab:scaling}
\begin{tabular}{lccc}
\toprule\toprule
System $(\sqrt{s_{\rm NN}})$ & Centrality & \multicolumn{2}{c}{$S_{f}$} \\
\cmidrule(lr){3-4}
 & & $p_{\rm T}/n_{q}$ & $(m_{\rm T} - m_{0})/n_{q}$ \\
\midrule
Au+Au ($200$ GeV) & $0$--$20\%$  & $0.075$ & $0.047$ \\
                  & $20$--$40\%$ & $0.086$ & $0.071$ \\
\midrule
                  & $10$--$20\%$ & $0.305$ & $0.151$ \\
Pb+Pb ($5.02$ TeV)& $30$--$40\%$ & $0.366$ & $0.188$ \\
                  & $60$--$70\%$ & $0.574$ & $0.424$ \\
\bottomrule\bottomrule
\end{tabular}
\end{table}

\subsubsection{NCQ scaling at $\sqrt{s_{\rm NN}} = 5.02$ $\rm TeV$}

The NCQ scaling analysis at LHC energy has previously been carried out by the ALICE Collaboration for Pb+Pb collisions at $\sqrt{s_{\rm NN}} = 5.02$ TeV~\cite{acharya2026alice}, though the goodness of the scaling was not rigorously established. Fig.~\ref{fig:NCQ_scaled_alice_data} presents the scaled distributions $v_{0}(p_{\rm T})/n_q$ as functions of $p_{\rm T}/n_q$ (\textit{top}) and $(m_{\rm T}-m_{0})/n_q$ (\textit{bottom}) for different centrality classes, compiled from the published ALICE data~\cite{acharya2026alice}. Only the statistical uncertainties are considered. A clear NCQ scaling is observed for \textit{central} (10--20\%) and \textit{semi-central} (30--40\%) collisions when the transverse kinetic energy per quark, $(m_{\rm T}-m_{0})/n_q$, is used as the scaling variable. In contrast, the scaling with $p_{\rm T}/n_q$ remains visibly weaker for the same centrality classes. For \textit{peripheral} (60--70\%) collisions, the scaling is found to break down for both variables, indicating a reduced degree of partonic collectivity. The violation of NCQ scaling in \textit{peripheral} events can be ascribed to the smaller system size and shorter lifetime of the produced medium, which limit the buildup of strong collective dynamics and possibly enhance the relative contribution from non-flow effects and initial-state fluctuations. Conversely, the improved universality of the $(m_{\rm T}-m_{0})/n_q$-scaled distributions across hadron species in \textit{central} collisions supports a scenario in which the observed collectivity is established at the partonic level.

To quantify the degree of NCQ scaling at LHC energies, we apply the same fitting and evaluation procedure defined in Sec.~\ref{subsub_sec:ncq_scaling_AuAu}, using the scaling parameter $S_f$ [Eq.~(\ref{eq:scaling_factor})]. The extracted $S_f$ values, summarized in Tab.~\ref{tab:scaling}, show that the scaling is systematically stronger for $(m_{\rm T}-m_{0})/n_q$ than for $p_{\rm T}/n_q$, and improves toward more \textit{central} collisions, while being substantially degraded in \textit{peripheral} events.
\section{Summary} \label{sec:5}

We present a systematic study of radial collectivity in Au+Au collisions at $\sqrt{s_{\rm NN}} = 200$~GeV using the differential observable $v_{0}(p_{\rm T})$ within the String Melting configuration of the \texttt{AMPT} model. The $v_{0}(p_{\rm T})$ exhibits a monotonic increase with transverse momentum, evolving from negative values at low-$p_{\rm T}$ to positive values at intermediate $p_{\rm T}$. The sign change occurs around $p_{\rm T} \approx 0.55$--$0.63$~GeV/$c$, indicating a characteristic redistribution of spectral yield induced by fluctuations of the radial expansion. The ratio $v_{0}(p_{\rm T})/v_{0}$ demonstrates centrality-independent scaling at low-$p_{\rm T}$, indicating a common dynamical origin of the spectral shape within the partonic stage. The weak dependence of $v_{0}(p_{\rm T})$ on pseudorapidity gap confirms the long-range nature of the correlations, and the observed factorization behavior further supports a collective origin. The $p_{\rm T}$-integrated observable $v_{0}$ decreases toward \textit{central} collisions and exhibits a sensitivity to the chosen reference momentum interval. However, the normalized quantity $v_{0}/v_{0}^{0\text{--}5\%}$ shows near-universal scaling across different $p_{\rm T}^{\rm ref}$ selections, indicating that $v_{0}$ captures event-by-event fluctuations of the global spectral shape independent of the specific momentum window. Similar collective signatures for inclusive charged hadrons, including long-range pseudorapidity correlations, factorization behavior, and centrality-independent scaling of $v_{0}(p_{\rm T})/v_{0}$, have been reported by the ATLAS Collaboration in Pb+Pb collisions at $\sqrt{s_{\rm NN}} = 5.02$~TeV~\cite{aad2026atlas}, indicating that the global features of radial collectivity observed at LHC energies are also present at RHIC.

For identified particles ($\pi^{\pm}, K^{\pm}, p + \overline{p}$), $v_{0}(p_{\rm T})$ exhibits mass-ordering at low-$p_{\rm T}$ and meson--baryon separation at intermediate-$p_{\rm T}$. In \textit{central} collisions, $v_{0}(p_{\rm T})/n_{q}$ follows robust \textit{Number of Constituent Quark} (NCQ) scaling with $\left(m_{\rm T} - m_{0}\right)/n_{q}$, consistent with partonic collectivity, while this scaling breaks down in \textit{peripheral} collisions, where the partonic phase of the system becomes increasingly short-lived. The scaling behavior is found to be more precise at $\sqrt{s_{\rm NN}} = 200$~GeV than at higher energies, in qualitative agreement with earlier $v_{2}$ measurements~\cite{singha2016}. 
These features are consistent with the observations reported by the ALICE Collaboration in Pb+Pb collisions at $\sqrt{s_{\rm NN}} = 5.02$ TeV~\cite{acharya2026alice}, supporting the interpretation that radial collectivity is predominantly established during the partonic stage and extending the paradigm of quark-level dynamics from anisotropic to isotropic flow.
\nocite{*}
\appendix
\section{Multiplicity dependence of $v_{0} \sqrt{\rm N_{ch}}$}

\begin{figure*}
    \centering
    \includegraphics[width=0.9\linewidth]{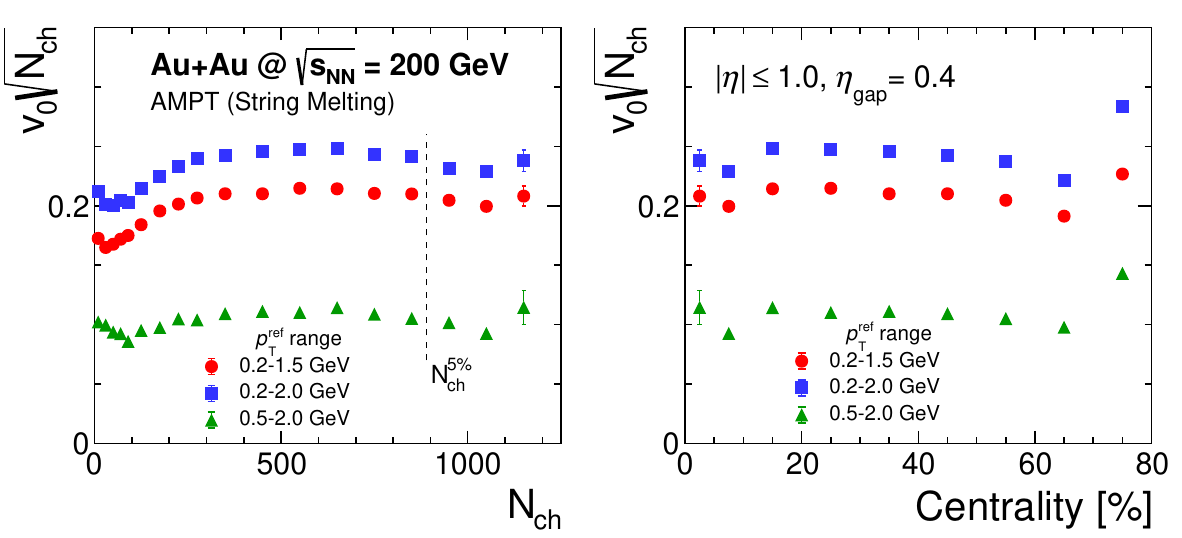}
    \caption{(Color online) $v_{0} \sqrt{\rm N_{ch}}$ versus $\rm N_{ch}$ (\textit{left}) and centrality (\textit{right}) for Au+Au collisions at $\sqrt{s_{\rm NN}} = 200$ GeV, shown for three $p_{\rm T}^{\rm ref}$ selections.}
    \label{fig:v0rootNch}
\end{figure*}

The relationship between radial flow fluctuations and system size is investigated by examining the product $v_{0} \sqrt{\rm N_{ch}}$ as a function of charged-particle multiplicity (and centrality) in Au+Au collisions at $\sqrt{s_{\rm NN}} = 200$ GeV. In Fig.~\ref{fig:v0}, it was seen that $v_{0}$ decreases with increasing multiplicity, approximately following a $1/\sqrt{\rm N_{ch}}$ scaling (verified by the power-law fit, Tab.~\ref{tab:v0_fit_parameters}). If this scaling holds, then multiplying $v_{0}$ by $\sqrt{\rm N_{ch}}$ should yield a quantity that remains relatively constant across different centrality classes. Fig.~\ref{fig:v0rootNch} demonstrates this behavior in our \texttt{AMPT-SM} simulations, showing that $v_{0} \sqrt{\rm N_{ch}}$ exhibits only a weak dependence on multiplicity across the centrality range studied. This approximate constancy confirms that the magnitude of global radial flow fluctuations is primarily governed by geometric fluctuations in the initial collision zone, which naturally scale with the inverse square root of the number of participating sources. The mild residual centrality dependence observed, particularly in \textit{peripheral} collisions, may reflect the increasing role of non-flow correlations or the limitations of the simple independent source picture in describing the full dynamics of the expanding system. These observations are consistent with recent measurements by the ATLAS Collaboration for Pb+Pb collisions~\cite{aad2026atlas}.
\section{Power law fit to the ATLAS $v_{0}$ data}

In this appendix, the ATLAS data~\cite{aad2026atlas} are replotted and fitted using a power-law. The resulting fit parameters are listed in Tab.~\ref{tab:v0_fit_parameters}.

\begin{figure}
    \centering
    \includegraphics[width=\linewidth]{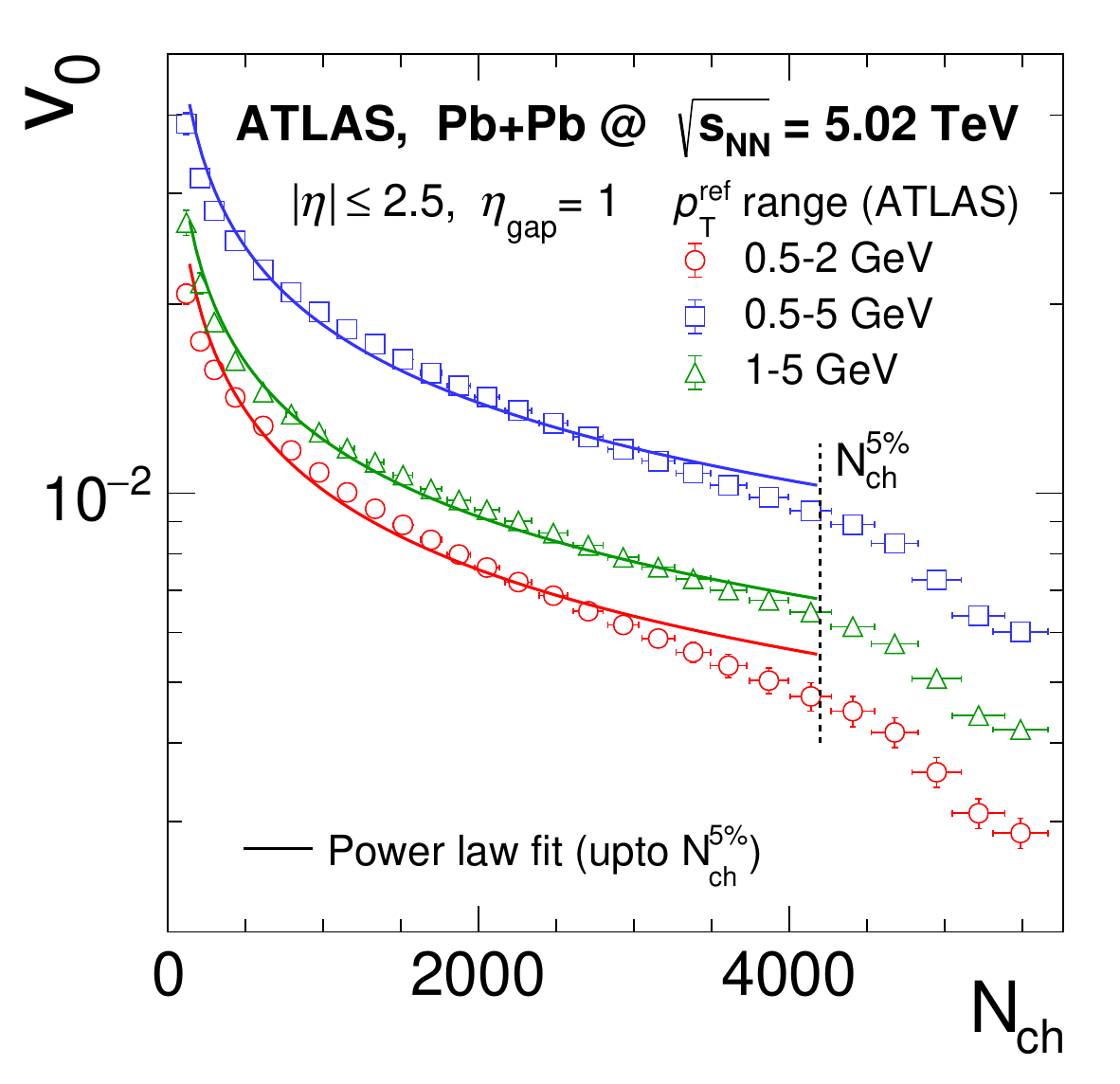}
    \caption{(Color online) The integrated radial flow measure $v_{0}$ as a function of $\rm N_{ch}$ for three $p_{\rm T}^{\rm ref}$ ranges for Pb+Pb collisions at $\sqrt{s_{\rm NN}} =$ 5.02 TeV. Power-law fits are applied to all $v_{0}$ results up to $\rm N_{ch}^{5\%}$. The figure is compiled using the experimental data of ATLAS~\cite{aad2026atlas}.}
    \label{fig:ATLAS}
\end{figure}
\bibliography{apssamp}
\end{document}